# Nematic state stabilized by off-site Coulomb interaction in iron-based superconductors

Xiao-Jun Zheng,[1] Zhong-Bing Huang,[2,3,*] Da-Yong Liu,[1] and Liang-Jian Zou[1,4,†]

[1]*Key Laboratory of Materials Physics, Institute of Solid State Physics,*
*Chinese Academy of Sciences, P. O. Box 1129, Hefei 230031, China*
[2]*Faculty of Physics and Electronic Technology, Hubei University, Wuhan 430062, China*
[3]*Beijing Computational Science Research Center, Beijing 100084, China*
[4]*Department of Physics, University of Science and Technology of China, Hefei 230026, China*
(Dated: July 8, 2014)

Using a variational Monte Carlo method, we investigate the nematic state in iron-base superconductors based on a three-band Hubbard model. Our results demonstrate that the nematic state, formed by introducing an anisotropic hopping order into the projected wave function, can arise in the underdoped regime when a realistic off-site Coulomb interaction $V$ is considered. We demonstrate that the off-site Coulomb interaction $V$, which is neglected so far in the analysis of iron-base superconductors, make a dominant contribution to the stabilization of nematic state. We calculate the doping dependencies of the anisotropic properties such as the unequal occupation of $d_{xz}$ and $d_{yz}$ orbitals, anisotropies of kinetic energy and spin correlations, and show that they are all suppressed upon electron doping, which are consistent with the intrinsic anisotropies observed by optical spectrum measurement and ARPES experiments.

Recently, electronic nematic phase, in which the discrete lattice rotational symmetry is broken but the translational symmetry is retained [1, 2], has been observed and widely discussed in correlated electron systems, such as bilayer Ruthenate [3], high-$T_c$ cuprates [4–10], and iron-based superconductors (FeSCs) [11–16]. Since its possible relation to the high-$T_c$ superconductivity, this phenomenon in the later two systems has been highly attached importance and received considerable attentions.

In high-$T_c$ cuprates, the nematic state is induced either by a $d$-wave Pomeranchuk instability [17–19] or via quantum melting of charge stripes [1, 20]. As for FeSCs, the situation can be quite different. Firstly, in contrast to the Mott insulator in cuprates, the parent phase of FeSCs is metallic, suggesting that the conduction electrons in FeSCs are less correlated. Secondly, in most of FeSCs, the ordering vector of the long-range magnetic order is $(\pi, 0)$ or $(0, \pi)$ [21–23], which is different from the $(\pi, \pi)$ ordering vector in high-$T_c$ cuprates. This stripe-like antiferromagnetic (SAFM) state which occurs at $T_N$ is always preceded by or coincident with a tetragonal-to-orthorhombic structural transition at $T_S$ (see [24] for review). Both of these two transitions break the fourfold rotational symmetry down to a twofold symmetry. Furthermore, it is widely believed that high-$T_c$ cuprates could be described by a single-band model, distinct from the multi-band electronic structure of FeSCs.

The nematic phase in FeSCs is manifested by the onset of anisotropies of dc resistivity [11], optical conductivity [25, 26] and orbital polarization of $d_{yz}$ and $d_{xz}$ Fe states [12] above $T_S$ in the tetragonal structure. These anisotropies in experiments are much stronger than those from slight difference of lattice constants driven by the structural transition. For this reason, many theorists consider that the nematic state are electronic correlation driven, and two scenarios are proposed: One scenario, considers magnetic fluctuations as the driving force of nematic state [15, 16, 27]; The other one [14, 28, 29] takes the orbital ordering as the driving force, i.e., the degeneracy between 3$d$ $xz$ and $yz$ orbitals is spontaneously broken, and the resulting orbital occupation renormalizes the exchange constants and triggers the magnetic transition at a lower temperature. Currently, it remains unclear how to distinguish these two phenomenological scenarios due to the coupling between spin and orbital degrees of freedom.

While quite a number of experimental phenomena have been explained by either of these two scenarios, an important issue needs to be clarified is that while electronic correlations are taken as the driving force of nematic state, most of analysis was based on phenomenological models, and no solid analysis based on realistic models has been done so far. In the mean field analysis [7] and the numerical calculation [10] based on the realistic three-band model for high-$T_c$ cuprate, substantial electron correlations including the on-site and off-site Coulomb interactions are needed to stabilized the nematic state. As electronic correlations in FeSCs are weaker than in high-$T_c$ cuprates, whether they can play the same role as the one in high-$T_c$ cuprates is under doubt.

To clarify the above issue and give a further insight into the nematic state in FeSCs, in this Letter, we perform a variational Monte Carlo study on a three-band (3B) Hubbard model with large lattice sizes ranging from 20×20 to 24×24. A highlight of our model is that an off-site Coulomb interaction $V$ is included, which is neglected so far, to our knowledge, in all the analysis of FeSCs, but played an important role in the formation of nematic order in $Sr_3Ru_2O_7$ [30] and high-$T_c$ cuprates [7, 10, 19, 31]. The numerical results presented below (see Figs. 1 and 3) indicate that similar to high-$T_c$ cuprates, $V$ is crucial for stabilizing the nematic state in FeSCs. Our variational calculations also confirm that the intrinsic anisotropies in the nematic are suppressed upon electron doping, just as the ones that observed by optical spectrum measurement and ARPES experiments.

The two-dimensional 3B Hubbard model is given as

$$H = H_0 + U_1 \sum_{i\alpha} n_{i\alpha\uparrow} n_{i\alpha\downarrow}$$
$$+ \sum_{i,\alpha<\beta,\sigma,\sigma'} \left[ (U_2 - J\delta_{\sigma\sigma'}) n_{i\alpha\sigma} n_{i\beta\sigma'} \right]$$
$$+ J \sum_{i,\alpha<\beta} \left( c^\dagger_{i\alpha\uparrow} c^\dagger_{i\beta\downarrow} c_{i\alpha\downarrow} c_{i\beta\uparrow} + c^\dagger_{i\alpha\uparrow} c^\dagger_{i\alpha\downarrow} c_{i\beta\downarrow} c_{i\beta\uparrow} + \text{H.c.} \right)$$
$$+ V \sum_{\langle ij \rangle} n_i n_j, \quad (1)$$

here $H_0$ is the kinetic part of the Hamiltonian, with transfer parameters $t^0_{\alpha\beta}[\Delta x, \Delta y]$ taken from Ref.[32]. We define $N$ as the number of sites and $n$ the average number of electrons per site. For the undoped case, $n=4$. The doping level $x$ is then defined as $x = n - 4$. The interaction part of the model includes intraorbital and interorbital Coulomb interaction $U_1$, $U_2$, the Hund coupling $J$, as well as the Coulomb interaction $V$ between nearest-neighbor (NN) sites.

The wave function we use is as following:

$$|\psi\rangle = P_G |\psi_{\text{MF}}\rangle = g_1^{\hat{N}_1} g_2^{\hat{N}_2} g_V^{\hat{N}_V} g_J^{\hat{N}_J} |\psi_{\text{MF}}\rangle, \quad (2)$$

where

$$\hat{N}_1 = \sum_{i,\alpha} n_{i\alpha\uparrow} n_{i\alpha\downarrow}, \hat{N}_2 = \sum_{i,\alpha<\beta} n_{i\alpha} n_{i\beta},$$
$$\hat{N}_V = \sum_{\langle ij \rangle} n_i n_j, \hat{N}_J = \sum_{i,\alpha<\beta,\sigma} n_{i\alpha\sigma} n_{i\beta\sigma}. \quad (3)$$

$g_1$, $g_2$, $g_J$ are the variational parameters controlling the number of electrons residing in the same and different on-site orbitals. $g_V$ controls the number of electrons on the NN sites.

To investigate the nematic state, an anisotropic hopping order (AHO) with order parameter $\delta_{\text{var}}$ is introduced. This kind of introducing nematic order has successfully captured the nature of the nematic state in high-$T_c$ cuprates [8, 10]. A non-interacting variational Hamiltonian $H_{\text{MF}}$ is then obtained by substituting some of hopping parameter $t^0_{\alpha\beta}[\Delta x, \Delta y]$ in $H_0$ by

$$t^{\text{MF}}_{\alpha\beta}[\Delta x, 0] = (1 + \delta_{var}) t^0_{\alpha\beta}[\Delta x, 0];$$
$$t^{\text{MF}}_{\alpha\beta}[0, \Delta y] = (1 - \delta_{var}) t^0_{\alpha\beta}[0, \Delta y]. \quad (4)$$

The wave function $|\psi_{\text{MF}}\rangle$ is then obtained by diagonalizing the quadratic Hamiltonian $H_{\text{MF}}$. It is well-known that the ground state of FeSCs is magnetic state at low doing. However, during our calculation, we do not include the magnetic order but take AHO as the only order parameter. This is because that the nematic state breaks the $C_4$ symmetry while keeping the spin O(3) symmetry [16], namely, it is defined in the paramagnetic regime before the magnetic long-range order set in. What we are interested is the microscopic origin of the unusual anisotropy in the paramagnetic regime with the tetragonal structure.

Unless otherwise stated, the values of the interacting parameters are $U_1=2.0$, $U_2=1.0$, $J=0.5$ and $V=0.5$ in unit of eV in this Letter, which correspond to typical values of iron-based superconductors. According to the *ab initio* calculations [33], we consider that setting $V=0.5$ is also appropriate.

Our calculations are performed for square lattices with periodic boundary conditions along the x and y directions. The ground state energy $\langle \psi | H | \psi \rangle$ is calculated using a standard Markovian chain Monte Carlo approach with Metropolis update algorithm, and is optimized with respect to the variational parameters. During the optimization, a quasi-Newton method combined with the fixed sampling method [34, 35] is used. In the figures presented below, the statistical errors are smaller than the symbol size unless otherwise stated.

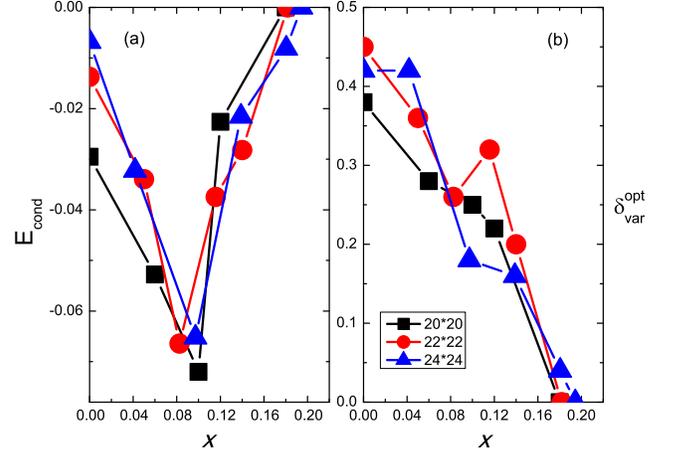

FIG. 1. (Color online) Condensation energy $E_{\text{cond}}$ (a) and optimized value of $\delta_{\text{var}}$ (b) as a function of electron doping on the 20×20, 22×22 and 24×24 lattices.

The condensation energy per unit cell $E_{\text{cond}}$ ($=\left[E\left(\delta^{\text{opt}}_{\text{var}}\right) - E(0)\right]/N$, with $\delta^{\text{opt}}_{\text{var}}$ being the optimized AHO parameter) as a function of doping is presented in Fig.1 (a). The results on the 20×20, 22×22 and 24×24 lattices consistently show that $|E_{\text{cond}}|$ exhibits a nonmonotonic doping dependence, with a maximum at finite doping around $x = 0.08 \sim 0.10$, and vanishes when $x$ is larger than 0.18. This behavior is quite different from that in high-$T_c$ cuprates, where a monotonic decrease of the condensation energy with increasing the doping density was observed [8, 10]. The condensation energies, with the largest value around 70meV, provide a strong evidence that the nematic state in FeSCs can be driven purely by electronic correlations. The doping dependence of $\delta^{\text{opt}}_{\text{var}}$ is shown in Fig.1 (b). One can see that $\delta^{\text{opt}}_{\text{var}}$ has a maximum value in the undoped case, and then is suppressed by increasing the electron doping. The error bar in Fig.1 (b) is about 0.05~0.08. This large error is due to the finite size effect [8, 10]. We will come back to this point later.

With the development of nematic order, the FSs spontaneously break their fourfold symmetry. As shown in the left panel of Fig. 2, the FSs at $x=0$ in the normal state are highly symmetric. In the nematic state, one of the main consequences is that all of the FSs become twofold symmetric, as seen in the right panel of Fig. 2. The electronic Fermi pockets around

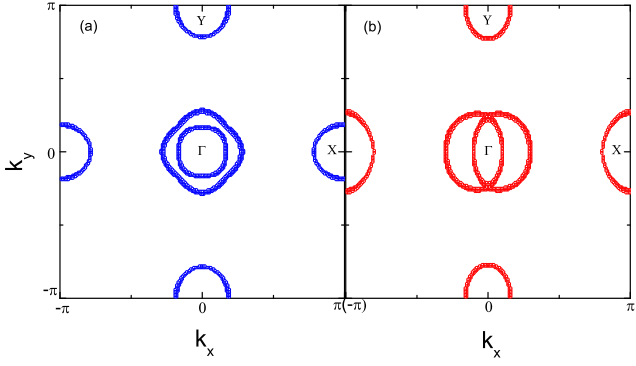

FIG. 2. (Color online) Fermi surfaces in the normal state ($\delta_{var}=0.0$) (a) and in the nematic state ($\delta_{var}=0.5$) (b) at $x=0$.

Y points considerably shrink along the x-direction, whereas the electronic Fermi pockets around X points expand along the y-direction; The hole FSs around Γ point display similar changes, and the rotational symmetry is reduced to a two-fold one. This kind of FS distortion was also observed in high-$T_c$ cuprates [10, 18, 19, 36].

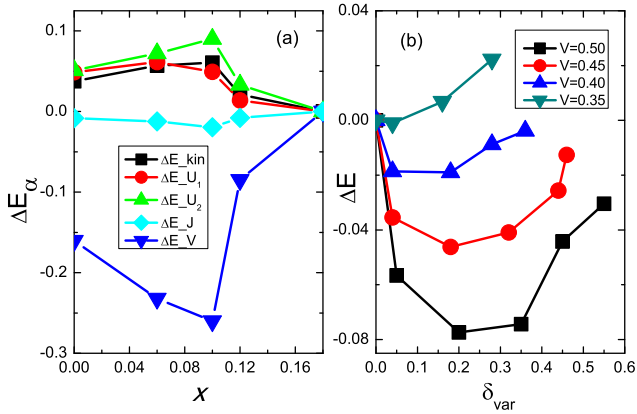

FIG. 3. (Color online) (a) Different energy contributions to the condensation energy as a function of electron doping on the 20×20 lattice; (b) Total energy change as a function of $\delta_{var}$ for different values of $V$ on the 20×20 lattice at $x = 0.10$. The different parts of Hamiltonian and the value of $V$ are indicated by the shape of symbol in (a) and (b), respectively.

In order to identify the physical origin for the formation of nematic phase, in Fig.3 (a) we present different energy contributions $\Delta E_\alpha = \left[E_\alpha\left(\delta_{var}^{opt}\right) - E_\alpha(0)\right]/N$ as a function of electron doing on the 20×20 lattice, with $\alpha$ representing different components of the Hamiltonian. In contrast to the positive contributions from the kinetic and on-site $U$ parts, a pronounced gain of Coulomb potential energy from the $V$ part demonstrates that the off-site Coulomb interaction $V$ plays a crucial role in stabilizing the nematic state. Fig. 3 (b) shows that with decreasing $V$, $E_{cond}$ is reduced dramatically and vanishes for $V \leq 0.35$. Considering that Coulomb screening effect will reduce the magnitude of $V$ as the electron doping is increased, it is expected that both $|E_{cond}|$ and $\delta_{var}^{opt}$ in the realistic electron-doped system should be smaller than those presented in Fig. 1 with a fixed $V$. For this reason, the extent of doping regime in which the nematic phase exists will shrink if a more realistic $V$ is used, making our results more comparable with the experiments, where the nematic characters were observed only in a narrow underdoped regime.

The role of $V$ in the stabilization of nematic state in FeSCs has not been investigated before. Whereas, the contribution of $V$ to the $d$-wave Pomeranchuk instability has been studied in the extended one-band Hubbard model. It has been demonstrated that the exchange part of $V$ in momentum space, which has the form of

$$V(\mathbf{k}) = -2V \sum_{\mathbf{k}'} \left[\cos\left(k_x - k'_x\right) + \cos\left(k_y - k'_y\right)\right] n(\mathbf{k}'), \quad (5)$$

enhance the $d$-wave compressibility of the FS and consequently leads to a FS distortion instability. phenomenological forward scattering model. We consider that in the FeSCs, the FSs of the normal state, which contain two equal electron pockets centered at the $X$ and $Y$ points as showed in Fig.2 (a), is unstable in the present of interaction $V$, sine the effective interaction between electrons around $(\pi, 0)$ and $(0, \pi)$ points in the Brillouin is positive according to Eq.5. While the FS is distorted with $\delta_{var} \neq 0$ as showed in Fig.2 (b), energy saved by reducing the positive electron-electron interaction between the two electron pockets and by enhancing the negative intra-pocket interaction around $X$. In this way, the nematic state is stabilized by $V$.

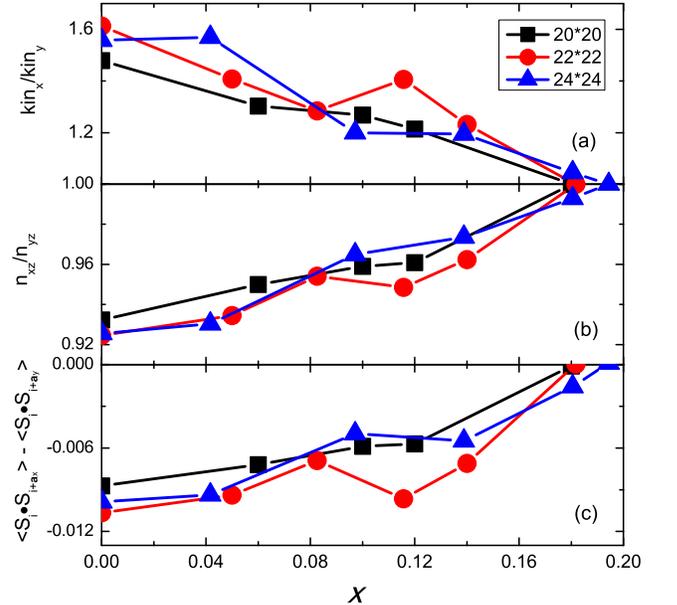

FIG. 4. (Color online) Ratio of $kin_x$ and $kin_y$ (a), $n_{xz}/n_{yz}$ (b) and $\langle S_i \cdot S_{i+\mathbf{a}_x}\rangle - \langle S_i \cdot S_{i+\mathbf{a}_y}\rangle$ (c) as a function of electron doping on the 20×20, 22×22 and 24×24 lattices. The kinetic energies $kin_x$ and $kin_y$ are defined in the text.

The doping dependence of nematic state is also an interesting issue. Figs. 4(a) and 4(b) show the ratios of kinetic energies along the x and y directions and of electron occupations in the $d_{xz}$ and $d_{yz}$ orbitals, which are defined as $\text{kin}_x/\text{kin}_y = \left\langle \sum_{i\alpha\beta\sigma\Delta x} t^0_{\alpha\beta}[\Delta x, 0] c^\dagger_{i,\alpha\sigma} c_{i+x,\beta\sigma} \right\rangle / \left\langle \sum_{i\alpha\beta\sigma\Delta y} t^0_{\alpha\beta}[0, \Delta y] c^\dagger_{i,\alpha\sigma} c_{i+y,\beta\sigma} \right\rangle$ and $n_{xz}/n_{yz}$, respectively. One can see that they decrease monotonically upon doping. The suppression of anisotropy of orbital occupation is in agreement with the behaviour of energy splitting of bands with dominant $xz$ and $yz$ characters observed by ARPES [12]. Figs. 4(c) displays the difference between spin correlations along the x and y directions, defined as $\langle S_i \cdot S_{i+\mathbf{a}_x} \rangle - \langle S_i \cdot S_{i+\mathbf{a}_y} \rangle$, with $\mathbf{a}_x$ and $\mathbf{a}_y$ denoting the unit vectors along the two vertical directions. The simultaneous anisotropies of orbital occupations and spin correlations demonstrate that in FeSCs the orbital and spin degrees of freedom are coupled together, which brings difficult to the disentangling of the orbital and magnetic scenarios. These anisotropies showed in Figs. 4, as with the order parameter $\delta^{\text{opt}}_{\text{var}}$ showed in Figs. 1 (b), are all suppressed upon doping. Our results, combining with the ARPES experimental [12] and optical spectrum measurement [38], confirm that the magnitude of intrinsic anisotropy is reduced upon electron doping. From this point of view, the anisotropy of dc resistivity, which becomes more pronounced with increasing Co doping [38–41], cannot be understand by the intrinsic electronic anisotropy alone. We suggest that a combination of the intrinsic nematicity with the anisotropic impurity scattering introduce by dopant Co might provide a comprehensive understanding of the dc anisotropy in FeSCs.

Another thing in Fig. 4 one needs to notice is that although the curves corresponding to different lattices basically exhibit the same trend, they are seen to be size dependent and show a non-monotonic behaviour at some doping levels. This is due to the finite size effect as we mentioned before. Due to the finite size of the lattices, the k-space is discrete. As a result, the FSs change in a discontinuous way when the value of the variational parameter $\delta_{\text{var}}$ changes. This brings uncertainty to $\delta^{\text{opt}}_{\text{var}}$ as showed in 1 (b), and consequently brings uncertainties to the properties showed in Figs. 4. It is these uncertainties, which different from lattice to lattice, make the curves to be size dependent and deviate from the overall trend at some doping levels. However, we consider that these errors do not change our conclusion that the presence of off-site interaction $V$ stabilizes the nematic phase. One can see that although the FSs are not exactly the same, the nematic instabilities obtained in all these three lattices are substantial , this can be see from fig.1 (a) that all the condensation energies obtained from different lattices have the substantial values at low doping.

In conclusion, we have demonstrated that the nematic state in FeSCs can be driven by electron correlations. Our results emphasize that the off-site Coulomb interaction $V$ between NN Fe ions plays an important role. We obtain the condensation energy $E_{cond}$ and the optimized order parameter $\delta^{\text{opt}}_{\text{var}}$ in the nematic state as functions of doping, and show that the suppression of $\delta^{\text{opt}}_{\text{var}}$ upon electron doping is consistent with the intrinsic anisotropies observed by optical spectrum measurement and ARPES experiments. We propose that the combination of intrinsic nematicity with anisotropic impurity scattering might provide a comprehensive understanding of the dc anisotropy in FeSCs.

This work was supported by the NSFC of China under Grant No. 11074257 and 11274310. Z.B.H. was supported by NSFC under Grant Nos. 11174072 and 91221103, and by SRFDP under Grant No.20104208110001. Numerical calculations were performed in Center for Computational Science of CASHIPS.